\documentclass[conference]{IEEEtran}
\pdfoutput=1
\IEEEoverridecommandlockouts
\usepackage{cite}
\usepackage{amsmath,amssymb,amsfonts}
\usepackage{algorithmic}
\usepackage{graphicx}
\usepackage{textcomp}
\usepackage{xcolor}
\usepackage{xspace}
\usepackage[export]{adjustbox} %
\usepackage{csquotes}
\usepackage{url}

\def\BibTeX{{\rm B\kern-.05em{\sc i\kern-.025em b}\kern-.08em
    T\kern-.1667em\lower.7ex\hbox{E}\kern-.125emX}}
\newcommand{\numRepos}{101\xspace}
\newcommand{\numTotalIssues}{53\,288\xspace}
\newcommand{\numTrainingsetIssues}{41\,771\xspace}
\newcommand{\numTestsetIssues}{11\,517\xspace}
\newcommand{\numValidationsetIssues}{100\xspace}

\newcommand{\numMarkdownDocumentationFiles}{5\,262\xspace}

\newcommand{\numMarkdownArtifactLines}{284\,267\xspace}
\newcommand{\numMarkdownTextLines}{219\,598\xspace}

\newcommand{\numTrainArtifactLines}{910\,478\xspace}
\newcommand{\numTrainTextLines}{360\,047\xspace}
\newcommand{\numTestArtifactLines}{193\,727\xspace}
\newcommand{\numTestTextLines}{36\,579\xspace}

\newcommand{\numBalancedTraininSet}{720\,094\xspace}
\newcommand{\numOFourTraininSet}{288\,038\xspace}
\newcommand{\numBalancedTestSet}{73\,158\xspace}

\newcommand{\numManualTestSet}{1\,816\xspace}
\newcommand{\numManualTestArtifactPortion}{76\,\%\xspace}

\newcommand{\numNlonDataSet}{6\,000\xspace}
\newcommand{\numNlonDataSetArtifactPortion}{29\,\%\xspace}

\newcommand{\numTenKLinesClassifiedInSeconds}{0.72\xspace}

\newcommand{\numManualTestSetInterraterCohen}{0.96\xspace}

\newcommand{\numNlonInterraterCohen}{0.88\xspace}
\newcommand{\numNlonInterraterFone}{0.94\xspace}

\newcommand{\numTestsetFOne}{0.96\xspace}
\newcommand{\numTestsetRocAuc}{0.96\xspace}

\newcommand{\numResearcherTwoFOne}{0.93\xspace}
\newcommand{\numResearcherTwoRocAuc}{0.95\xspace}

\newcommand{\numPretrainedOnNlonFabioRocAuc}{0.85\xspace}

\newcommand{\numNlonOnResearcherTwoRocAuc}{0.83\xspace}

\newcommand{\numBootstrapNlonDataLowerFOne}{0.91\xspace}
\newcommand{\numBootstrapNlonDataUpperFOne}{0.94\xspace}
\newcommand{\numBootstrapNlonDataMeanFOne}{0.93\xspace}

\newcommand{\numBootstrapNlonDataMeanRocAuc}{0.93\xspace}

\newcommand{\numNlonTrainedOnOurDataResearcherTwoScoreFOne}{0.90\xspace}
\newcommand{\numNlonTrainedOnOurDataResearcherTwoScoreRocAuc}{0.92\xspace}

\newcommand{\numRegexLinewiseValidationSetTwoFOne}{0.65\xspace}
\newcommand{\numRegexLinewiseValidationSetTwoRocAuc}{0.77\xspace}
\newcommand{\numRegexTicketwiseValidationSetTwoFOne}{0.85\xspace}
\newcommand{\numRegexTicketwiseValidationSetTwoRocAuc}{0.91\xspace}

\usepackage{balance}

\usepackage[bookmarks=false]{hyperref}
\hypersetup{nolinks=false}

\usepackage{fancyhdr}
\begin{document}

\title{Identifying non-natural language artifacts in bug reports}

\author{\IEEEauthorblockN{Thomas Hirsch}
\IEEEauthorblockA{
\textit{Graz University of Technology}\\
Graz, Austria \\
thirsch@ist.tugraz.at}
\and
\IEEEauthorblockN{Birgit Hofer}
\IEEEauthorblockA{
\textit{Graz University of Technology}\\
Graz, Austria \\
bhofer@ist.tugraz.at}
}

\maketitle

\begin{abstract}
Bug reports are a popular target for natural language processing (NLP).
However, bug reports often contain artifacts such as code snippets, log outputs and stack traces.
These artifacts not only inflate the bug reports with noise, but often constitute a real problem for the NLP approach at hand and have to be removed.
In this paper, we present a machine learning based approach to classify content into natural language and artifacts at line level implemented in Python.
We show how data from GitHub issue trackers can be used for automated training set generation, and present a custom preprocessing approach for bug reports.
Our model scores at \numResearcherTwoRocAuc ROC-AUC and \numResearcherTwoFOne F1 against our manually annotated validation set, and classifies 10k lines in \numTenKLinesClassifiedInSeconds seconds.
We cross evaluated our model against a foreign dataset and a foreign R model for the same task. %
The Python implementation of our model and our datasets are made publicly available under an open source license.
\end{abstract}

\begin{IEEEkeywords}
NLP, bug reports, data cleaning, artifact removal
\end{IEEEkeywords}

\thispagestyle{fancyplain}
\footnotetext{\textcopyright 2021 IEEE.  Personal use of this material is permitted.  Permission from IEEE must be obtained for all other uses, in any current or future media, including reprinting/republishing this material for advertising or promotional purposes, creating new collective works, for resale or redistribution to servers or lists, or reuse of any copyrighted component of this work in other works.}

\section{Introduction}

Natural language processing (NLP) approaches analyzing documents originating from software development processes are an increasingly popular and promising research field.
In particular, bug reports are very prominent targets for NLP approaches~\cite{Zhang2015}.
NLP applications on the basis of textual bug reports are used, for example, to
categorize the impact and root causes of bug reports~\cite{Zhou2021},
to classify bugs automatically according to the ODC classification scheme~\cite{Thung2012}, 
to assign programmers to bug reports~\cite{Mani2019,Devaiya2021}, 
to locate the source code that needs to be changed to fix a bug~\cite{Zhou2012},
to label the severity of a bug~\cite{Kumar2021,Kukkar2019},
to prioritize bugs~\cite{Ortu2016}, 
to detect duplicates~\cite{Kukkar2020}, 
to distinguish bug reports from other issues~\cite{Chawla2015},
and to find security related bug reports~\cite{Goseva-Popstojanova2018}.

However, in contrast to classic NLP problems, bug reports are often cluttered with non-natural language artifacts such as code snippets, stack traces, log outputs, config files, and file listings.
While some approaches detect and leverage specific artifacts---most notably information retrieval (IR) approaches \cite{Saha2013}---other NLP tasks consider such artifacts as mere noise that may even have negative effects on the task's performance, e.g., language detection and automated root cause classification approaches.
The size of such artifacts also can pose a problem for some approaches and technologies, as especially un-shortened stack traces, core dumps, and thread dumps lead to often unwanted growth of vocabulary, and excessive runtimes.
For example, we found bug reports as big as 200\,kB of uncompressed text\footnote{See e.g. \url{https://github.com/redisson/redisson/issues/2291}} when we created our dataset.

Issue trackers usually provide formatting mechanisms, such as Markdown, to allow users to format their issue comments.
One of the simplest forms of artifact detection would be to parse issue descriptions for fragments formatted as code blocks.
Unfortunately, not all users/reporters use them properly.\footnote{See e.g. \url{https://github.com/haraldk/TwelveMonkeys/issues/37}} %
Therefore, relying on such formatting alone is not a viable option for artifact detection.

Since certain artifacts add valuable information to some approaches (e.g., stack traces for automated fault localization),
researchers developed numerous techniques for identifying and parsing specific artifact types.
Several researchers \cite{Tan2014a,Ray2014,Soltani2020BugReports} manually built sets of regular expressions after they had investigated the underlying dataset.
Although labor intensive, this approach works reasonably well for datasets originating from a small number of software projects.

However, regular expressions have to be adapted when using them on other software projects, and manually annotating data is time-consuming, as pointed out by M\"antyl\"a \textit{et al.}~\cite{Mantyla2018}.
Manually created rule sets do not scale to bigger datasets due to the size and number of required regular expressions necessary to account for different logging frameworks, code style guidelines, built systems, configuration file formats, underlying OSs, and IDEs.
These scalability and portability issues led researchers to the application of machine learning (ML) techniques~\cite{Mantyla2018,Bacchelli2012}.
While ML approaches circumvent the manual creation of rules, they introduce the need for manually annotated training sets.

In this work, we propose and discuss a pragmatic approach that minimizes manual labor required when creating an ML classifier.
Our approach does not require extensive knowledge about the artifacts that are supposed to be removed, while providing good classification performance at a low computational cost once trained.
To achieve this, we tackle aforementioned problems from two sides:
(1)~by automating the training set creation process, and 
(2)~by using a custom preprocessing step. %
The automated training set creation process leverages project documentation Markdown files, and GitHub Markdown annotated bug reports.
In our custom preprocessing step, white spaces and special characters alongside other features are tokenized and therefore become parts of the feature vectors used by the ML-based classifier algorithm.

We use a standard Python ML library as basis of our implementation, and we provide an easy to use pretrained model that can be used in preprocessing for various NLP tasks concerning textual bug reports.
We train and validate our approach on \numRepos Java software projects hosted on GitHub.
Our model's performance on a manually created validation set is \numResearcherTwoRocAuc ROC-AUC and \numResearcherTwoFOne F1.
The model can be trained in less than two minutes, and classifies 10k lines in \numTenKLinesClassifiedInSeconds seconds.
We test our approach on a foreign dataset and discuss its portability and limitations.

This work is structured as follows:
We discuss related work in Section~\ref{sec:relatedwork} and define the problem  in Section~\ref{sec:problemdefinition}.
Section~\ref{sec:approach} contains a detailed description of our approach.
In Section~\ref{sec:evaluation}, we present our results, discussion of these results, limitations, and threats to validity.
We conclude this paper in Section~\ref{sec:conclusion}.

\section{Related work}\label{sec:relatedwork}

InfoZilla~\cite{Bettenburg2008} detects stack traces, source code, patches, and enumerations in bug reports using regular expressions, island parsing and heuristics.
Bacchelli \textit{et al.}~\cite{Bacchelli2011} used island parsing to extract structured data from natural language documents such as emails.
Later on, they proposed to use supervised ML to classify the content of emails line-by-line into natural language, junk, code, patch and stack trace.
To train and test the classifier, they manually classified the content of nearly 1500~emails from four software systems~\cite{Bacchelli2012}.

Rigby and Robillard \cite{Rigby2013} developed an automated code element resolution tool (ASE) that is based on island parsing.
ASE automatically extracts code elements %
from natural language documents such as StackOverflow posts and determines which elements are important for the document.
Ye \textit{et al.}~\cite{Ye2017} use a semi-supervised ML approach to detect API mentions in text written on social platforms.

Ponzanelli \textit{et al.}~\cite{Ponzanelli2015} provide a parsed dataset from Stack Overflow that contains heterogeneous abstract syntax trees for Java code, stack traces, XML/HTML elements and JSON fragments.
They used island parsing to identify these fragments.

Calefato \textit{et al.}~\cite{Calefato2019} reported that they tried to use regular expressions to remove code snippets from email text, but found this approach to not scale well enough---in particular when several programming languages are used.
This highlights the need for more generic approaches for artifact detections such as machine learning.

The work that is closest to ours is the Natural Language or Not (NLoN) Package \cite{Mantyla2018}.
This R package classifies text lines into text or artifact by using eleven language features and character tri-grams.

\section{Problem definition}\label{sec:problemdefinition}
While intuition tells us that the line between natural language or non-natural language should be a clear cut, closer investigation reveals the complexity of this problem and a certain amount of overlap of the two domains.
Examples of such border cases are code comments and bug report templates:
Comments contained in code snippets are natural language texts.
However, they may not have been authored by the bug reporter.
Bug report templates include headers, questions, and other text.
While these are natural language, they are again not written by the bug reporter and can be considered automatically generated text.
Migration from other bug tracking systems often introduces generated text portions.
These are also natural language, but their origin is artificial.
As far as we are aware, there exists no formal definition, established guideline, or agreement within the research community working with textual bug reports on what is to be considered natural language when dealing with bug reports.

For this work, we define artifacts and natural language portions of bug reports as follows:
We consider text that was typed by the bug reporter as natural language, and content that was copy-pasted from an IDE, terminal, or other tool to be an artifact.
Automatically generated natural language text of the bug tracker, template, or migration processes are considered natural language.
Comments in pasted code snippets, elaborate natural language logging messages and error messages are considered artifacts.
Further, we consider standalone URLs and Markdown links as artifacts.

Occurrences of non-natural language portions in a natural language sentence are limited mostly to variable names, class names, and short formulas or mathematical equations.
Removing such occurrences may render a natural language sentence syntactically and semantically incorrect and unreadable for a human.
We therefore consider a line of natural language text interweaved with non-natural language portions as natural language.
Log outputs or code snippets always start on a new line.
Thus, we detect artifacts on a line by line level and describe the task as a binary classification problem similar to M\"antyl\"a \textit{et al.}~\cite{Mantyla2018}.

\section{Approach}\label{sec:approach}
First, we explain in Section~\ref{sec:features} what features are used in our machine learning approach.
In Section~\ref{sec:trainingset}, we discuss the creation of the training, test and validation sets, before we discuss the setup of the ML approach in Section \ref{sec:mlapproach}.

\subsection{Feature Selection}\label{sec:features}

For humans, separating artifacts from natural language is often possible without the need of actually reading the text.
Formatting and structure plays a significant role in a human's capability to classify a given text segment very fast.
For example, indentation of code snippets provides a very good indicator.
Therefore, we will include representations of whitespaces in the feature vectors used by the ML classifier.

A closer look at artifacts further reveals that frequency and position of special characters also carry a significant amount of information for our task.
While the most common special characters in English text are \enquote*{,} and \enquote*{.}, the characters \enquote*{$<$}, \enquote*{$>$}, and \enquote*{/} are probably the most common in XML.
For this reason, we tokenize special characters to include them in the feature vectors.
The full replacement table can be found in the online appendix; an excerpt of this table is shown in Table~\ref{tab:specialchartokens}.

\begin{table}[tbp]
\caption{Excerpt of introduced tokens}
  \label{tab:specialchartokens}
\begin{center}
\begin{tabular}{ll}
\hline
\textbf{Character / Regex}                            & \textbf{Token}    \\
\hline
\textvisiblespace\ \textvisiblespace\ \textit{(two whitespaces)} & Jdoublespace      \\
\textbackslash{}t                      &  Jtabulator           \\
(                                      &  Jroundbracketopen    \\
\{                                     &  Jcurlybracketopen    \\
;                                      &  Jsemicolon           \\
\hline
\verb!([A-Z]?[a-z0-9]+)([A-Z][a-z0-9]*)+!     &  Jcamelcased          \\
\verb![0-9]+!                                 &  Jnumber              \\
\hline
\end{tabular}
\end{center}
\end{table}

Regarding the position of special characters, lines of English text will often end with \enquote*{.}, \enquote*{?}, and \enquote*{!}, while lines of Java code will often end with \enquote*{\{}, \enquote*{\}} or \enquote*{;}.
A bag of words (unigram) approach is not suitable to encapsulate such position information.
Thus, we add tokens that represent the beginning and end of a line, and employ tri-gram vectorization.

\subsection{Training, Test and Validation Sets}\label{sec:trainingset}
For our experiments, we mined \numRepos open source Java projects hosted on GitHub.
All of the projects utilize GitHub's built-in issue tracker.
We collected \numTotalIssues issue tickets that were labeled with \enquote*{bug}, \enquote*{defect}, or \enquote*{regression}.
Furthermore, we used the projects' documentation as additional source of training data. 

\begin{figure}[htbp]
	\centering                       %
		\includegraphics[width=1.0\columnwidth,trim=1.3cm 5.1cm 3.9cm 0.7cm, clip]{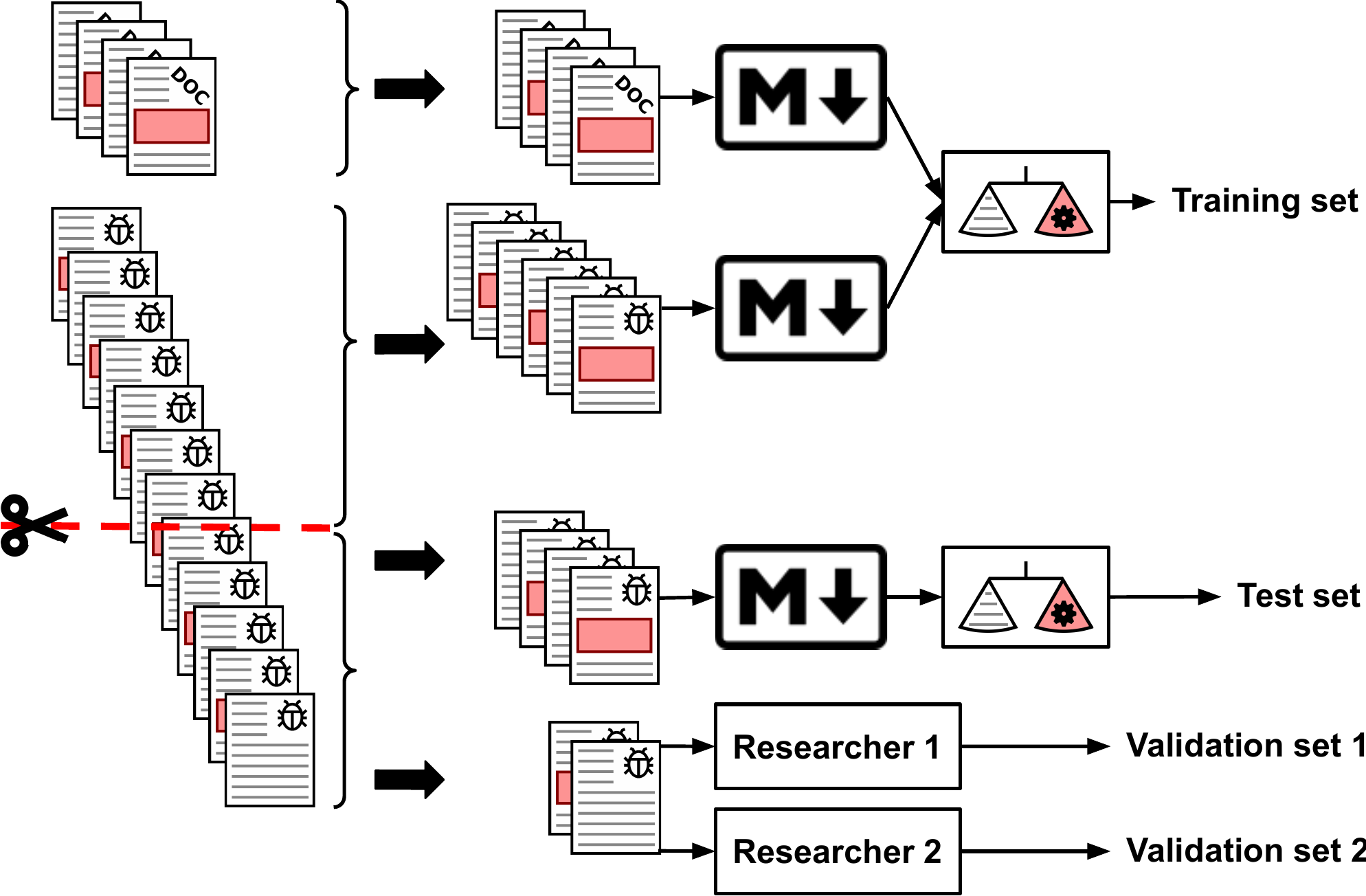}
	\caption{Training, test and validation set creation}
	\label{fig:ProcessingBugReports}
\end{figure}

Figure~\ref{fig:ProcessingBugReports} illustrates the training, test and validation set creation process.
We divided the set of issue tickets into training and test portions (see Section~\ref{sec:testTrainingSplit} for details).
The larger set with \numTrainingsetIssues~issue tickets is used together with \numMarkdownDocumentationFiles project documentation files as basis for the training set.
Details on the processing of the issue tickets and project documentation files are provided in Sections~\ref{sec:issues} and \ref{sec:doc}.
The smaller set with \numTestsetIssues issue tickets is used to source the test set.
From these issue tickets, \numValidationsetIssues tickets are manually inspected by the authors of this paper to provide validation sets (see Section~\ref{sec:validation}).
Furthermore, we reused the NLoN dataset as additional validation set (see Section~\ref{sec:nlonDataset}).

\subsubsection{Test and training split}\label{sec:testTrainingSplit}
Since parts of our dataset are used in other research, we  separate the training and test set along this line:
The test set exclusively consists of issue tickets that have commits linked to them %
as they are used in downstream research that will utilize this pretrained classifier model as preprocessing step.

The training set contains all remaining issue tickets, i.e., %
 \numTrainArtifactLines lines of non-natural language text, and \numTrainTextLines lines of natural language text.
The test set contains \numTestArtifactLines, and \numTestTextLines lines accordingly.
We balance both sets to contain an equal number of lines of artifacts and natural language text.
The resulting balanced training set contains \numBalancedTraininSet lines, the resulting test set contains \numBalancedTestSet lines in total.

\subsubsection{Issue tickets}\label{sec:issues}
GitHub's built-in issue tracker offers Markdown\footnote{https://guides.github.com/features/mastering-markdown/} for reporters to format their issue reports.
For this work, we focus mainly on the following Markdown features:
Triple ticks that start and end a code highlighting block, indentation by four spaces signaling a code block, lines that are full quotes from start to end, Markdown style links, tables, URLs, and embedded images.
If all issue reporters would make use of these Markdown features to properly wrap non-natural language artifacts, the task of artifact removal would be trivial, but this is unfortunately not the case.

About 45\,\% of the issue tickets contain Markdown formatted code blocks.
From these issue tickets, we manually examined 300~issue tickets 
and found that  93.3\,\% used Markdown formatting consistently.
The other issue tickets contained code snippets, stack traces and/or log output that were not properly wrapped Markdown blocks.
Analogue to the above, we also sampled 300~issue tickets from the 55\,\% that do not contain Markdown code blocks.
18\,\% of those tickets contain artifacts that are not properly Markdown formatted.

\begin{figure}[tbp]
	\centering                       %
		\includegraphics[width=1.0\columnwidth,trim=0.9cm 6.1cm 9.0cm 4.9cm, clip]{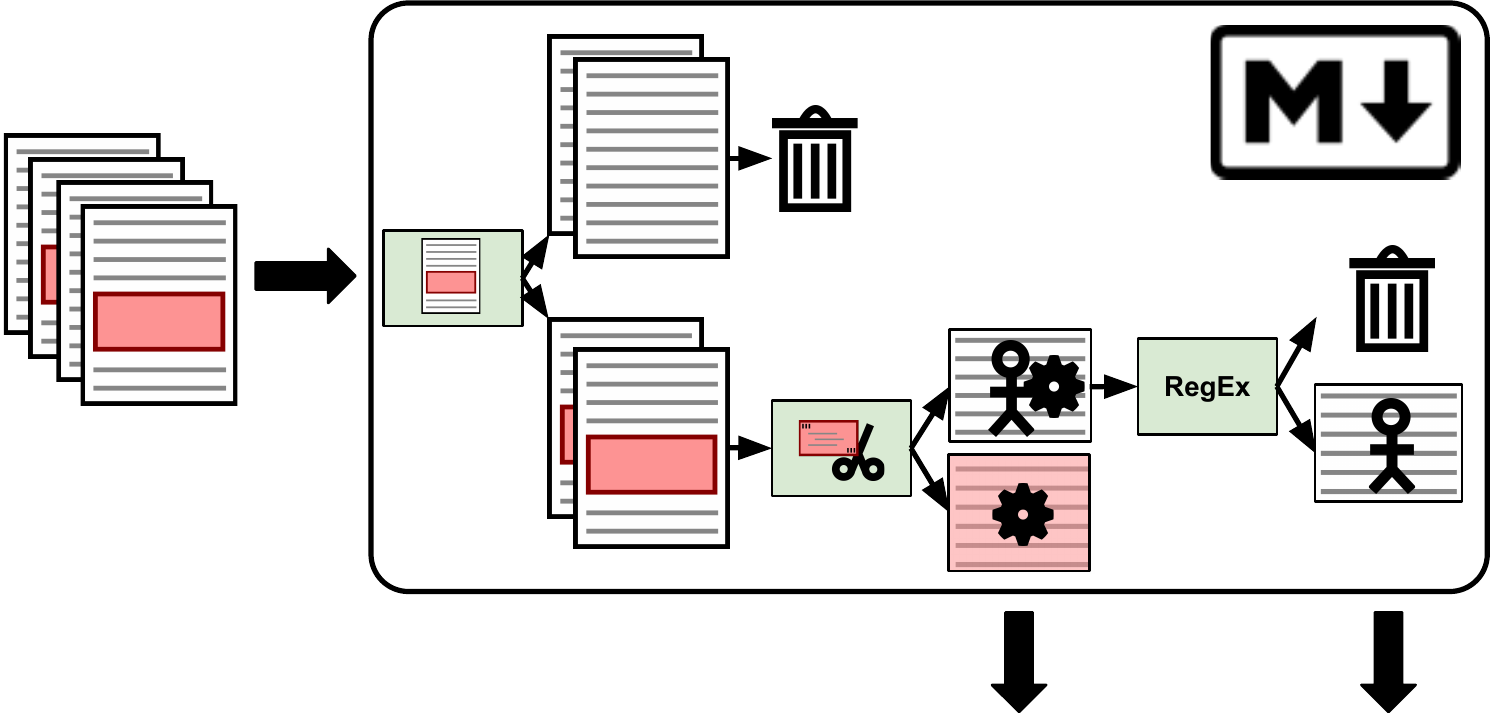}
	\caption{Automatic separation of human-written text and artifacts}
	\label{fig:Trainingset}
\end{figure}

Since Markdown is not consistently used, the trivial approach of using Markdown features to identify artifacts is insufficient.
However, we can leverage the issue tickets that contain Markdown highlighted code highlighting features to automate the creation of a data set for an ML classifier.
Figure~\ref{fig:Trainingset} illustrates this process.

In detail, we search for all issue tickets that contain such Markdown code block tags, and split these texts accordingly into natural language and non-natural language portions.
To capture these various Markdown code blocks, we use a small set of six regular expressions.
This process is based on the assumption that if a reporter utilizes Markdown for one portion of his/her report, that he/she will also do this for other portions of the same report.

However, this assumption does not always hold, and therefore produces supposedly natural language text portions that in fact are artifacts of some kind.
We examine the produced datasets, and augment the above described approach by employing a set of regular expressions to filter common artifact types from the natural language side of the dataset.
The first part of these regular expressions can be easily reused in any context: 
Two regular expressions remove Unix and Windows style prompts, and three regular expressions remove json and xml like content.
The second part of regular expressions depend on the used programming language:
Five regular expressions  specifically aim at Java code, and four regular expressions  target logging formats.
We finally use two regular expressions to remove lines whose formatting does not allow to distinguish them via regular expressions (e.g. Markdown block quotes using \enquote*{$>$} that are used for reply or followup in conversations as well as for code highlighting).

To measure the noise in the final dataset, Researcher 1 manually inspected  600 artifact lines and 600 natural language lines randomly sampled from the \emph{test set}.
While the artifact portion of the generated dataset did not contain any natural language lines, %
the natural language text portion contains 35 lines that are artifacts.
Most of these mislabeled lines constitute corner cases, e.g., information on operating systems including detailed version numbers.
We estimate that about 6\,\% of the lines labeled as natural language and therefore 3\,\% of all lines contained in our \emph{training set} and \emph{test set} are mislabeled.

\subsubsection{Project documentation files}\label{sec:doc}
The documentation files of a software project are a great source of training data for the purpose of artifact detection.
Given the documentation in the form of Markdown files, we employ the same approach as for issue tickets discussed above (see Figure~\ref{fig:Trainingset}).
We processed \numMarkdownDocumentationFiles \emph{.md} files, resulting in \numMarkdownArtifactLines artifact lines and \numMarkdownTextLines lines of natural language.
The resulting dataset obtained from documentation files is significantly cleaner since documentation files Markdown features are usually used properly and uniformly by the projects' maintainers.

\subsubsection{Validation set}\label{sec:validation}
To estimate the effect of aforementioned noise on the classifier model and to enable reproducible and objective performance measurements, we created a human annotated validation set to serve as ground truth.
We randomly sampled 100~issue tickets from the set of issue tickets that were used to create the test set.
Both authors individually classified these bug reports, annotating each line in the bug report to be either \emph{artifact} or \emph{natural language}.
They were provided with the full textual bug reports for this work, in order to provide context and therefore a qualitatively better ground truth.
The Cohen's Kappa~\cite{Cohen1960} inter-rater agreement of Researcher~1 and Researcher 2 is \numManualTestSetInterraterCohen and can be interpreted as \enquote*{almost perfect} agreement~\cite{Landis1977}.
We noticed that Researcher~1 tended to classify lines in the gray area as artifacts, while Researcher~2 classified them as natural text. 
The resulting data sets, excluding empty lines, contain \numManualTestSet lines.
\numManualTestArtifactPortion of the lines are artifacts, the remainder are natural language.
Thus the data sets are imbalanced.

Since Researcher 1 implemented the preprocessing pipeline, we consider Researcher 2 as the ground truth for performance measure as to avoid any unwanted bias.
From hereon we refer to these datasets as \emph{validation set~1} and \emph{validation set~2}, corresponding to their human classifiers Researcher~1 and Researcher~2.

\subsubsection{NLoN data set}\label{sec:nlonDataset}
The dataset of M\"antyl\"a \textit{et al.}~\cite{Mantyla2018} is publicly available\footnote{https://github.com/M3SOulu/NLoN}.
It was built from three different sources: comments on Mozilla's issue tracker of multiple different projects (C++), chats from Kubernetes' public slack channel (Go), and messages on Apache Lucene's  mailing list archives (Java).
For each source, 2000~data samples were manually labeled as natural text or artifact.
The full dataset comprises \numNlonDataSet lines, of which \numNlonDataSetArtifactPortion are artifacts.

\subsection{Machine Learning}\label{sec:mlapproach}
Here we present our model's setup in detail, from preprocessing, through utilized ML algorithms, to our model evaluation strategy.

\subsubsection{Preprocessing}
In the first step, we use regular expressions and basic string operations to perform the replacements and tokenizations discussed in Section~\ref{sec:features}.
This step is implemented as a scikit-learn transformer.
Doing so enables us to utilize standard tokenization and vectorization steps of the Python library, without any adaptations.
We do not replace stop words, as these are important features for our task to differentiate natural language from other artifacts.
Further, we do not perform case folding, as this also carries some information for the task at hand (e.g. all caps words are more common in artifacts).
To encapsulate positional information of the tokens in the feature vectors (as discussed in Section~\ref{sec:features}), we vectorize into uni-, bi-, and tri-grams that are combined into a single feature vector using a simple count vectorizer.

\subsubsection{ML models}
We use classic ML models as Support Vector Machines (SVM), Random Forrest Classifier (RFC), Logistic Regression Classifier (LRC), and Multinomial Naive Bayes (MNB), due to their ease of use and little requirements in terms of computational resources for training and prediction.
We do not perform hyperparameter tuning, and keep the default values of the classifiers in the used library (MNB: $\textit{alpha}=1.0$, SVM: $C=1.0$, RFC: $\textit{nEstimators}=100$, LRC: $C=1.0$).
In a preliminary experiment, the classification performance and capabilities of all classifiers were very similar, but the prediction and training times varied.
Given the similarity in classification performances, we chose  SVM  for the following experiments.

\subsubsection{Performance evaluation}
We apply the test and training sets presented in Section~\ref{sec:trainingset}.
We measure the classification performance on the test set and the validation set and the prediction runtime on the test set.
To enable comparison and a discussion of the external validity, we measure the performance of our model trained on our dataset on the NLoN~\cite{Mantyla2018} dataset, as well as the performance of our model when trained on the NLoN dataset.

\section{Evaluation}\label{sec:evaluation}
First, we present the results in Section~\ref{sec:results} before we discuss them in Section~\ref{sec:discussion}.
Then, we deal with the limitations and threats to validity in Section~\ref{sec:threats}.
\subsection{Results}\label{sec:results}

\begin{figure}[t]
  \centering
  \includegraphics[width=\linewidth]{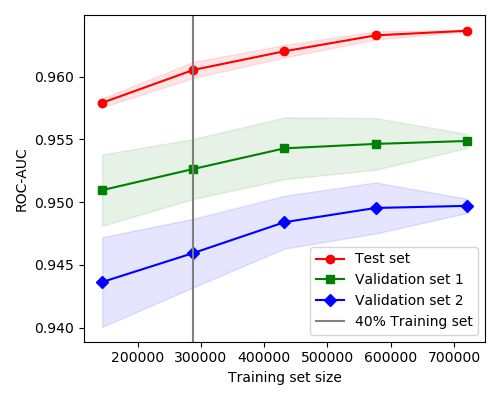}
  \caption{Mean ROC-AUC learning curve and standard deviation}
      \label{fig:learningCurveF1SVM}
\end{figure}

Figure~\ref{fig:learningCurveF1SVM} shows the mean performance and standard deviation of ROC-AUC performance for ten randomized runs in relation to different training set sizes.
Our model's performance on both \emph{validation sets} continually increases with the \emph{training set} size.
However, as the classification performance increases, so does the time required for training the model, and more importantly, the memory and storage space requirements of the pretrained model.

We therefore randomly sample 40\,\% of the \emph{training set} to be used to train our final model, to control these properties.
This resulting training set contains \numOFourTraininSet lines, 
and the results reported in this section stem from a model trained on this set.
This final model can be trained in less than two minutes, its size is under 60MB, and it takes on average \numTenKLinesClassifiedInSeconds seconds to classify 10k lines of input (as single-threaded Python 3.7 process on Debian 10, on Intel i5-4460, 3.20GHz).
The first two rows of Table~\ref{tab:performanceXvalidationCombined} contain the average classification performance from ten runs of this model as macro average F1 and ROC-AUC.

\begin{table}[htbp]
\caption{Evaluation results of our model and NLoN model}
  \label{tab:performanceXvalidationCombined}
\begin{center}
\begin{tabular}{lllcc}
\hline
\textbf{Model} & \textbf{Training}   & \textbf{Evaluation} & \textbf{F1} & \textbf{ROC-AUC} \\
\hline
\multicolumn{5}{l}{\textit{Base validation}} \\
Ours & 40\%~Training set & Test set                     & 0.96                 & 0.96       \\
Ours & 40\%~Training set & Validation set~2             & 0.93                 & 0.95       \\
\hline
\multicolumn{5}{l}{\textit{Cross validation}}	\\	
Ours           & 40\%~Training set   & NLoN dataset          & 0.86                 & 0.85       \\
NLoN           & NLoN dataset        & Validation set~2      & 0.81                 & 0.83       \\
Ours           & NLoN dataset        & NLoN dataset          & 0.93                 & 0.93       \\
NLoN           & 40\%~Training set   & Validation set~2      & 0.90                 & 0.92       \\
\hline
\end{tabular}
\end{center}
\end{table}

We cross evaluated our pretrained model against a pretrained NLoN model.
Each model was used to classify the data from the other model, NLon model on validation set 2, and our model on NLoN-Researcher~2 targets.
The results of this evaluation are given in the third and fourth row of Table~\ref{tab:performanceXvalidationCombined}.

We further cross evaluated our classifier approach against the NLoN approach by training and validating the models on the opposing datasets.
An NLoN classifier was trained on our training set and scored against our validation set.
Analog to this, we trained and scored our classifier on the NLoN dataset.
Since our employed SVM algorithm is sensitive to imbalanced training sets, we balanced the NLoN dataset by downsampling the majority class (natural language).
Given the small size of the dataset we used the Bootstrap algorithm with $alpha=0.95$ and $n=100$ to calculate the 95\,\% confidence intervals with 0.8/0.2 training/test splits.

The bottom two rows of Table~\ref{tab:performanceXvalidationCombined} show the resulting performance scores for both models.
Both, the F1 and ROC-AUC confidence intervals of our classifier on the NLoN dataset are  \numBootstrapNlonDataLowerFOne to \numBootstrapNlonDataUpperFOne with a mean of \numBootstrapNlonDataMeanFOne over all 100~iterations.
The NLoN classifier performed at F1 \numNlonTrainedOnOurDataResearcherTwoScoreFOne and ROC-AUC \numNlonTrainedOnOurDataResearcherTwoScoreRocAuc on our validation set.
Due to excessive training times for this NLoN classifier (multiple hours) we only performed a single training and evaluation run.

An important question is how well the automatic separation of text and artifacts with regular expressions as described in Section~\ref{sec:issues} performs on validation set 2.
Line-wise classification performs poorly with an F1 score of \numRegexLinewiseValidationSetTwoFOne and a ROC-AUC score of \numRegexLinewiseValidationSetTwoRocAuc.
Performing this task on complete and continuous bug reports aids this approach by enabling detection of Markdown triple quote code blocks and performs at \numRegexTicketwiseValidationSetTwoFOne F1 and \numRegexTicketwiseValidationSetTwoRocAuc ROC-AUC.

\subsection{Discussion}\label{sec:discussion}
Despite the noise in our automatically generated training and test sets, our model scores well.
The classification performance on our manually created validation set is close to the performance on our automatically generated test set with \numTestsetFOne vs. \numResearcherTwoFOne F1 scores and \numTestsetRocAuc vs. \numResearcherTwoRocAuc ROC-AUC scores.
We investigated misclassifications of our automated dataset generation process in comparison to our manual classification efforts and found that there is an overlap in the disagreement:
The same type of artifact misidentified by our automated dataset generation is also often mismatched between the two reviewers.
This together with the high classification performance of the model supports our assumption that our automated training and test set creation approach is valid.

The cross evaluation of our model and the NLoN model highlights the limitations in portability of such pretrained models.
The origins of the data (bug reports, comments on bug reports, mailing lists, or chats) seem to be an important factor.
Further, the different programming languages in the projects used to create these datasets heavily influence the type and form of artifacts found in these texts.

To investigate the reason for this low portability, we retrained and evaluated our classification model on the NLoN dataset.
The resulting classifier performed at an average of \numBootstrapNlonDataMeanFOne F1 and ROC-AUC.
Given an inter-rater F1 score of \numNlonInterraterFone of the NLoN dataset, these performance scores support the validity of our preprocessing and modeling approach.
These scores demonstrate that the low cross evaluation scores of the pretrained model arise from dissimilarities in the datasets, while the classifier pipeline including our custom preprocessing is well portable.

However, these scores are lower than the ones reported for our dataset.
The reasons for this are twofold:
First, and most importantly, the training set size plays a significant role to our approach as already shown in Figure~\ref{fig:learningCurveF1SVM}, and the NLoN dataset is significantly smaller than the training set used in our experiments.
Second, the inter-rater agreement within the NLoN dataset with a Cohen's Kappa of \numNlonInterraterCohen is lower than for our manual validation set (\numManualTestSetInterraterCohen).
This lower inter-rater agreement in the NLoN dataset can be explained by their process of randomly sampling lines for manual classification, while our manual classification process provided full bug tickets as context to the human classifier.

The performance comparison of our ML approach with simple  regular expression parsing on the validation set (F1 score \numResearcherTwoFOne for ML vs. \numRegexLinewiseValidationSetTwoFOne for regular expressions on a per line basis / \numRegexTicketwiseValidationSetTwoFOne for regular expressions and Markdown blocks) shows the advantage of ML based classifier models over simple regular expressions.
\subsection{Limitations and Threats to Validity}\label{sec:threats}
We used only open source projects in our experiments.
Therefore, we cannot generalize our results to closed source software projects.
Since a dataset containing an ample amount of complete bug reports from commercial software is hard to come by, we cannot validate our approach in this domain.

A threat to external validity is that our internal dataset was mined solely from GitHub.
Other bug trackers may encourage or discourage certain behaviors of reporters, may enforce more or less rules for more structured bug reports, and may vary in length, format, and tone of natural language.
Further, our internal dataset was constructed exclusively from Java projects.
Other programming languages may differ in frequency and usage of special characters and may employ different formatting rules, both being important features that we leverage in our approach.

We addressed this threat by validating our approach on the NLoN dataset, that was sourced from mailing lists, chat messages, and comments on bug reports. %
Two of the three projects used in the NLoN dataset are not written in Java (Go, C++).
Using this dataset, we showed that our preprocessing and ML classifier approach does not overfit on any features predominate in GitHub bug reports or language features, and is portable to other data sources.

Another threat to external validity is that the natural language portions used in our experiment are in English language. %
While logographic writing systems or other languages using non-Latin alphabets probably ease the task of distinguishing natural language from artifacts, our main concern are languages using a Latin alphabet.
Since our approach leverages the projects' documentation to mine natural language items, this poses an issue if the documentation and the bug reports are written in different languages.

A threat to internal validity is our manual classification effort in order to create a validation set.
It is subject to human error, and given the task at hand, also subject to human preference regarding what is actually considered an artifact or human language.
To counter this threat, two researchers independently classified the bug reports for the validation set.
We analyzed the inter-rater agreement on this dataset, and used the classifications of Researcher~2 who was not involved in the implementation of the system to prevent any bias based on implementation details of the approach.

\section{Conclusion}\label{sec:conclusion}
In this work, we present our Python framework for removing non-natural language text portions from bug reports.
This includes a process for automated training set generation, preprocessing steps aimed at feature extraction for this task, and an LSVM classifier trained on our data set.
We demonstrate the viability of this approach on a manually annotated dataset, by showing that despite the inherent noise in automatically created datasets we can achieve a F1 score of \numResearcherTwoFOne and a ROC-AUC score of \numResearcherTwoRocAuc.

The proposed preprocessing and machine learning portion leverages differences of natural language and artifacts as stack traces, code snippets, and similar that are normally lost in common NLP preprocessing approaches.
That being formatting and structure, frequency and position of special characters, and specific word constructs, e.g., camelcased names and names utilizing underscores.

M\"antyl\"a \textit{et al.}~\cite{Mantyla2018} manually investigated the underlying datasets to create a list of specific features (e.g. line ends with \enquote*{\{}) scoring exceptionally well in their evaluation.
Our proposed approach aims at enabling the ML algorithm to learn such features from the training set instead of explicitly providing them, by including aforementioned formatting and structure as well as special characters in tokenization and vectorization.
In doing so, we skip the manual effort required for feature identification and implementation.

We cross evaluated NLoN and our model by applying the pretrained models on the opposing validation/test sets.
Both pretrained models are limited in their portability, with ROC-AUC scores between \numNlonOnResearcherTwoRocAuc and \numPretrainedOnNlonFabioRocAuc.
However, %
we show that without any parameter tuning the model can be trained on a foreign dataset with reasonable performance.
We trained and scored our model on the NLoN dataset to enable comparison.
This yielded mean scores of \numBootstrapNlonDataLowerFOne F1 and \numBootstrapNlonDataMeanRocAuc ROC-AUC despite different programming languages used in the dataset and its origins of mailing lists, comments on bug reports, and chat messages, in contrast to our dataset of bug reports.
We found that our approach benefits from bigger datasets.
Such large datasets can be automatically created with our approach.

Our automated dataset creation process is based on parts of the input data being properly Markdown formatted.
We demonstrate our process on the basis of bug reports for Java projects mined from GitHub issue trackers.
We are confident that our approach can be %
easily ported to other programming languages, given that the underlying bug tickets are Markdown annotated. %
Depending on the noisiness of the new data, this may require some effort to add domain specific regular expressions.
We hope that our proposed approach is useful for researchers dealing with textual bug reports.

\section{Data availability}\label{sec:material}
The created datasets from \numRepos open source Java projects, the manual validation sets, and our model's Python source code are made publicly available on Zenodo\footnote{\url{https://doi.org/10.5281/zenodo.5519503}} and GitHub\footnote{\url{https://github.com/AmadeusBugProject/artifact_detection}}.

\section*{Acknowledgment}
The work described in this paper has been funded by the Austrian Science Fund (FWF): P 32653-N (Automated Debugging in Use).

\balance

\bibliographystyle{IEEEtran}
\bibliography{library}

\begin{thebibliography}{10}
\providecommand{\url}[1]{#1}
\csname url@samestyle\endcsname
\providecommand{\newblock}{\relax}
\providecommand{\bibinfo}[2]{#2}
\providecommand{\BIBentrySTDinterwordspacing}{\spaceskip=0pt\relax}
\providecommand{\BIBentryALTinterwordstretchfactor}{4}
\providecommand{\BIBentryALTinterwordspacing}{\spaceskip=\fontdimen2\font plus
\BIBentryALTinterwordstretchfactor\fontdimen3\font minus
  \fontdimen4\font\relax}
\providecommand{\BIBforeignlanguage}[2]{{%
\expandafter\ifx\csname l@#1\endcsname\relax
\typeout{** WARNING: IEEEtran.bst: No hyphenation pattern has been}%
\typeout{** loaded for the language `#1'. Using the pattern for}%
\typeout{** the default language instead.}%
\else
\language=\csname l@#1\endcsname
\fi
#2}}
\providecommand{\BIBdecl}{\relax}
\BIBdecl

\bibitem{Zhang2015}
J.~Zhang, X.~Y. Wang, D.~Hao, B.~Xie, L.~Zhang, and H.~Mei, ``{A survey on
  bug-report analysis},'' \emph{Science China Information Sciences}, vol.~58,
  pp. 1--24, feb 2015.

\bibitem{Zhou2021}
\BIBentryALTinterwordspacing
C.~Zhou, B.~Li, X.~Sun, and L.~Bo, ``{Why and what happened? Aiding bug
  comprehension with automated category and causal link identification},''
  \emph{Empirical Software Engineering}, vol.~26, no.~6, pp. 1--36, aug 2021.
  [Online]. Available:
  \url{https://link.springer.com/article/10.1007/s10664-021-10010-8}
\BIBentrySTDinterwordspacing

\bibitem{Thung2012}
F.~Thung, D.~Lo, and L.~Jiang, ``{Automatic defect categorization},'' in
  \emph{Working Conference on Reverse Engineering (WCRE)}, 2012, pp. 205--214.

\bibitem{Mani2019}
\BIBentryALTinterwordspacing
S.~Mani, A.~Sankaran, and R.~Aralikatte, ``{DeepTriage: Exploring the
  effectiveness of deep learning for bug triaging},'' in \emph{ACM India Joint
  International Conference on Data Science and Management of Data (CoDS-COMAD
  '19)}, jan 2019, pp. 171--179. [Online]. Available:
  \url{http://dl.acm.org/citation.cfm?doid=3297001.3297023}
\BIBentrySTDinterwordspacing

\bibitem{Devaiya2021}
\BIBentryALTinterwordspacing
D.~Devaiya, J.~Anvik, M.~Bheree, and F.~{Yeasmin Omee}, ``{Evaluating a Tool
  for Creating Bug Report Assignment Recommenders},'' in \emph{33rd
  International Conference on Software Engineering {\&} Knowledge Engineering},
  2021. [Online]. Available: \url{https://bugzilla.mozilla.org}
\BIBentrySTDinterwordspacing

\bibitem{Zhou2012}
J.~Zhou, H.~Zhang, and D.~Lo, ``{Where should the bugs be fixed? More accurate
  information retrieval-based bug localization based on bug reports},'' in
  \emph{International Conference on Software Engineering (ICSE 2012)}, 2012,
  pp. 14--24.

\bibitem{Kumar2021}
\BIBentryALTinterwordspacing
L.~Kumar, T.~G. Dastidar, L.~B. {Murthy Neti}, S.~M. Satapathy, S.~Misra,
  V.~Kocher, and S.~Padmanabhuni, ``{Deep-Learning Approach with DeepXplore for
  Software Defect Severity Level Prediction},'' in \emph{International
  Conference on Computational Science and Its Applications (ICCSA 2021)}.\hskip
  1em plus 0.5em minus 0.4em\relax Springer, Cham, sep 2021, pp. 398--410.
  [Online]. Available:
  \url{https://link.springer.com/chapter/10.1007/978-3-030-87007-2{\_}28}
\BIBentrySTDinterwordspacing

\bibitem{Kukkar2019}
\BIBentryALTinterwordspacing
A.~Kukkar, R.~Mohana, A.~Nayyar, J.~Kim, B.-G. Kang, and N.~Chilamkurti, ``{A
  Novel Deep-Learning-Based Bug Severity Classification Technique Using
  Convolutional Neural Networks and Random Forest with Boosting},''
  \emph{Sensors}, vol.~19, no.~13, pp. 2964:1--22, jul 2019. [Online].
  Available: \url{https://www.mdpi.com/1424-8220/19/13/2964/htm
  https://www.mdpi.com/1424-8220/19/13/2964}
\BIBentrySTDinterwordspacing

\bibitem{Ortu2016}
M.~Ortu, G.~Destefanis, S.~Swift, and M.~Marchesi, ``{Measuring high and low
  priority defects on traditional and mobile open source software},'' in
  \emph{7th International Workshop on Emerging Trends in Software Metrics
  (WETSoM 2016)}, may 2016, pp. 1--7.

\bibitem{Kukkar2020}
A.~Kukkar, R.~Mohana, Y.~Kumar, A.~Nayyar, M.~Bilal, and K.~S. Kwak,
  ``{Duplicate Bug Report Detection and Classification System Based on Deep
  Learning Technique},'' \emph{IEEE Access}, vol.~8, pp. 200\,749--200\,763,
  2020.

\bibitem{Chawla2015}
\BIBentryALTinterwordspacing
I.~Chawla and S.~K. Singh, ``{An automated approach for bug categorization
  using fuzzy logic},'' in \emph{8th India Software Engineering Conference
  (ISEC 2015)}.\hskip 1em plus 0.5em minus 0.4em\relax ACM, feb 2015, pp.
  90--99. [Online]. Available:
  \url{http://dl.acm.org/citation.cfm?doid=2723742.2723751}
\BIBentrySTDinterwordspacing

\bibitem{Goseva-Popstojanova2018}
K.~Goseva-Popstojanova and J.~Tyo, ``{Identification of Security related Bug
  Reports via Text Mining using Supervised and Unsupervised Classification},''
  in \emph{IEEE International Conference on Software Quality, Reliability and
  Security Identification (QRS'18)}, 2018, pp. 344--355.

\bibitem{Saha2013}
R.~K. Saha, M.~Lease, S.~Khurshid, and D.~E. Perry, ``{Improving bug
  localization using structured information retrieval},'' in \emph{28th
  IEEE/ACM International Conference on Automated Software Engineering (ASE
  2013)}, 2013, pp. 345--355.

\bibitem{Tan2014a}
L.~Tan, C.~Liu, Z.~Li, X.~Wang, Y.~Zhou, and C.~Zhai, ``{Bug characteristics in
  open source software},'' \emph{Empirical Software Engineering}, vol.~19,
  no.~6, pp. 1665--1705, oct 2014.

\bibitem{Ray2014}
B.~Ray, D.~Posnett, V.~Filkov, and P.~Devanbu, ``{A large scale study of
  programming languages and code quality in GitHub},'' in \emph{ACM SIGSOFT
  Symposium on the Foundations of Software Engineering (FSE'14)}, nov 2014, pp.
  155--165.

\bibitem{Soltani2020BugReports}
M.~Soltani, F.~Hermans, and T.~B{\"{a}}ck, ``{The significance of bug report
  elements},'' \emph{Empirical Software Engineering}, vol.~25, no.~6, pp.
  5255--5294, sep 2020.

\bibitem{Mantyla2018}
M.~M{\"{a}}ntyl{\"{a}}, F.~Calefato, and M.~Claes, ``{Natural Language or Not
  (NLoN) - A Package for Software Engineering Text Analysis Pipeline},'' in
  \emph{IEEE/ACM 15th International Conference on Mining Software Repositories
  (MSR)}, 2018, pp. 387--391.

\bibitem{Bacchelli2012}
A.~Bacchelli, T.~D. Sasso, M.~D'Ambros, and M.~Lanza, ``{Content classification
  of development emails},'' in \emph{34th International Conference on Software
  Engineering (ICSE 2012)}, 2012, pp. 375--385.

\bibitem{Bettenburg2008}
N.~Bettenburg, T.~Zimmermann, R.~Premraj, and S.~Kim, ``{Extracting structural
  information from bug reports},'' in \emph{International Conference on
  Software Engineering (ICSE 2008)}, 2008, pp. 27--30.

\bibitem{Bacchelli2011}
\BIBentryALTinterwordspacing
A.~Bacchelli, A.~Cleve, M.~Lanza, and A.~Mocci, ``{Extracting structured data
  from natural language documents with island parsing},'' in \emph{26th
  IEEE/ACM International Conference on Automated Software Engineering (ASE
  2011)}, 2011, pp. 476--479. [Online]. Available:
  \url{https://ieeexplore.ieee.org/abstract/document/6100103/}
\BIBentrySTDinterwordspacing

\bibitem{Rigby2013}
P.~C. Rigby and M.~P. Robillard, ``{Discovering essential code elements in
  informal documentation},'' in \emph{35th International Conference on Software
  Engineering (ICSE 2013)}, 2013, pp. 832--841.

\bibitem{Ye2017}
D.~Ye, Z.~Xing, C.~Y. Foo, J.~Li, and N.~Kapre, ``{Learning to extract API
  mentions from informal natural language discussions},'' in \emph{IEEE
  International Conference on Software Maintenance and Evolution (ICSME 2016)},
  jan 2017, pp. 389--399.

\bibitem{Ponzanelli2015}
L.~Ponzanelli, A.~Mocci, and M.~Lanza, ``{StORMeD: Stack overflow ready made
  data},'' in \emph{IEEE International Working Conference on Mining Software
  Repositories}, vol. 2015-Augus, aug 2015, pp. 474--477.

\bibitem{Calefato2019}
F.~Calefato, F.~Lanubile, and B.~Vasilescu, ``{A large-scale, in-depth analysis
  of developers' personalities in the Apache ecosystem},'' \emph{Information
  and Software Technology}, vol. 114, pp. 1--20, oct 2019.

\bibitem{Cohen1960}
\BIBentryALTinterwordspacing
J.~Cohen, ``{A Coefficient of Agreement for Nominal Scales},''
  \emph{Educational and Psychological Measurement}, vol.~20, no.~1, pp. 37--46,
  apr 1960. [Online]. Available:
  \url{http://journals.sagepub.com/doi/10.1177/001316446002000104}
\BIBentrySTDinterwordspacing

\bibitem{Landis1977}
J.~R. Landis and G.~G. Koch, ``{The Measurement of Observer Agreement for
  Categorical Data},'' \emph{Biometrics}, vol.~33, no.~1, p. 159, mar 1977.

\end{thebibliography}

\end{document}